\documentclass[fleqn,usenatbib]{mnras}

\usepackage{newtxtext,newtxmath}

\usepackage[T1]{fontenc}
\usepackage{ae,aecompl}

\usepackage{graphicx}	
\usepackage{amsmath}	
\usepackage{amssymb}	

\title[Star-formation in early-type galaxies]{Star-formation in CALIFA early-type galaxies. A matter of discs}

\author[J. M\'endez-Abreu et al.]{
J. M\'endez-Abreu,$^{1,2}$\thanks{E-mail: jairomendezabreu@gmail.com} S. F. S\'anchez,$^{3}$ A. de Lorenzo-C\'aceres$^{1,2}$
\\
$^{1}$Instituto de Astrof\'isica de Canarias, Calle V\'ia L\'actea s/n, E-38205 La Laguna, Tenerife, Spain\\
$^{2}$Departamento de Astrof\'isica, Universidad de La Laguna, E-38200 La Laguna, Tenerife, Spain\\
$^{3}$Instituto de Astronom\'ia, Universidad Nacional Aut\'onoma de M\'exico, A.P. 70-264, 04510 M\'exico, D.F., M\'exico.\\
}

\date{Accepted XXX. Received YYY; in original form ZZZ}

\pubyear{2015}

\begin{document}
\label{firstpage}
\pagerange{\pageref{firstpage}--\pageref{lastpage}}
\maketitle

\begin{abstract}
The star formation main sequence (SFMS) is a tight relation between the galaxy star formation rate (SFR) and its total stellar mass ($M_\star$). Early-type galaxies (ETGs) are often considered as low-SFR outliers of this relation. We study, for the first time, the separated distribution in the SFR vs. $M_\star$ of bulges and discs of 49 ETGs from the CALIFA survey. This is achieved using {\sc c2d}, a new code to perform spectro-photometric decompositions of integral field  spectroscopy datacubes. Our results reflect that: i) star formation always occurs in the disc component and not in bulges; ii) star-forming discs in our ETGs are compatible with the SFMS defined by star forming galaxies at $z \sim 0$; iii) the star formation is not confined to the outskirts of discs, but it is present at all radii (even where the bulge dominates the light); iv) for a given mass, bulges exhibit lower sSFR than discs at all radii; and v) we do not find a deficit of molecular gas in bulges with respect to discs for a given mass in our ETGs. We speculate our results favour a morphological quenching scenario for ETGs.
\end{abstract}




\section{Introduction}

Early-type galaxies (ETGs), a category encompassing lenticular and elliptical galaxies, are generally described as red, passive/retired, and morphologically featureless systems at the end of galaxy evolution. However, recent studies have demonstrated that ETGs are not such as simple systems, but they represent a whole family showing a wide range of specific angular momenta \citep{emsellem11, cappellari11}, bulge-to-total luminosity ratios \citep[$B/T$;][]{laurikainen11,kormendybender12,mendezabreu18}, cold and ionised-gas amounts \citep{sarzi10,davis11,serra12}, and star-formation modes \citep[][]{shapiro10}.

Despite most of the stellar mass in ETGs is formed at high redshift \citep{cowie96,thomas10,sanchez19}, the study of the recent star formation in these systems has provided new clues on their mass assembly. Studies of the ultra-violet (UV) emission in ETGs show that residual star formation is occurring in the outer parts of a non-negligible fraction of these galaxies \citep{jeong07,salimrich10,moffett12}, generally following distinctive spiral-like patterns noticeable in the ionised-gas \citep{gomes16b}, that have been also detected in atomic Hydrogen \citep{yildiz15}. Similarly, observations of molecular gas (mostly CO) have also proved its presence in ETGs and showed signatures of different modes of star formation \citep{shapiro10, young11}. Spitzer and GALEX observations have revealed that up to 10\% of the current stellar mass in ETGs is due to recent SF \citep{kaviraj07,schawinski07}. 

Still, the processes of triggering and shut-down of the star formation in ETGs are not well understood. The low efficiency of the star formation in these galaxies has been attributed to both external or internal mechanisms. The former are related to either major mergers where the gas is rapidly consumed in the violent starburst \citep{hopkins09} or processes related to high density environments such as gas stripping \citep{feldmann10}. Internal processes quenching the star formation can be related to energy feedback from either active galactic nuclei (AGNs) or star formation that heats the gas \citep{cattaneo09,schawinski09}, but also to disc stabilisation due to the presence of a massive spheroid at the galaxy center,  the so-called {\it morphological quenching} \citep{martig09,martig13}. Detailed studies of low-level extended star formation in early-type galaxies can therefore provide critical constraints on the relative importance of each of these mechanisms in regulating the mass growth in these systems. 

A powerful tool to  study the evolution of the star formation rate (SFR) in galaxies is provided by scaling relations such as the star formation main sequence (hereafter, SFMS), a tight relation between a galaxy's SFR and its total stellar mass \citep[$M_{\star}$;][]{brinchmann04,salim07}. This relationship exists out to high redshifts, increasing its normalization to higher values at earlier epochs such that SFRs at a fixed stellar mass are higher by a factor of $\sim$20 at $z \sim 2 $ \citep[e.g.,][]{daddi07,speagle14,sanchez19}. The position of the galaxies with respect to the SFMS, at any given epoch, give us information about whether the SFR is enhanced or suppressed relative to the {\it mean} for their stellar mass. Despite the large amount of literature in the topic, most of the works on the properties of the SFMS have been limited to studies of star-forming galaxies. Nevertheless, it is known that ETGs remain well below the SFMS \citep{noeske07,schiminovich07,sanchez19}, with various processes invoked as responsible for the differences in SFR with respect to star forming galaxies. These include: i) internal structure \citep[such as bars and bulges; e.g.][]{wuyts11,bluck14}, cold gas fraction \citep{saintonge12,tacconi13}, interaction with other galaxies \citep{scottkaviraj14,willett15} and the presence of an AGN \citep{ellison16, sanchez18}.

In this paper, we explore for the first time the SFR vs. $M_{\star}$ relation of ETGs separated into their bulge and disc components. To this aim, we use the new algorithm {\sc c2d} \citep{mendezabreu19} designed to perform spectro-photometric decompositions of galaxy structures using integral field spectroscopy (IFS), in combination with {\sc Pipe3D} \citep{sanchez16b}, a taylor-made pipeline to retrieve the stellar populations and ionised-gas properties from IFS. The application of {\sc c2d}+{\sc Pipe3D} to the sample of ETGs present in the CALIFA survey \citep{sanchez12} allows us to explore the SFR using the {\it fossil approach} \citep{gonzalezdelgado16,sanchez19}. Therefore allowing for a robust comparison throughout the whole sample independently of our ability to detect emission lines.

\section{CALIFA sample of ETGs}
\label{sec:sample}

The sample of ETGs with photometric discs analysed in this paper is extracted from the CALIFA data release 3 \citep[DR3;][]{sanchez16}. This last release comprises 667 galaxies covering a wide range of stellar masses and Hubble types. \citet{mendezabreu17} carried out a multicomponent multiband photometric decomposition of 404 galaxies present in the CALIFA DR3 using SDSS imaging. This includes all galaxies, but those with high inclination (i > 70 degrees) or in interaction \cite[see][for more details]{mendezabreu17}. From this subsample, 127 galaxies were visually classified as elliptical or lenticular (S0) galaxies in \citet{walcher14}. A further, purely photometric, analysis on the nature of these ETGs was performed in \citet{mendezabreu18} combining both a logical filtering and a statistical diagnostic to separate ellipticals (pure bulge) and lenticulars. We found that 41, 50, and 36 galaxies can be classified as pure-bulge (elliptical), lenticular, and unknown, respectively. The latter group implies that our photometric approach is not able to classify a given galaxy, and that either a simple S\'ersic or a S\'ersic+Exponential fit returns the same statistical solution. From the sample of 50 photometrically confirmed S0s, we discarded 29 galaxies showing a stellar bar at their centres, while the 36 unknown galaxies were analysed using their two-component (bulge and disc) best fit. After the {\sc c2d} fitting was performed, we discarded other 8 (4 lenticular and 4 unknown) galaxies because the code did not converge. Therefore, the final sample used in this paper comprises 49 ETGs analysed with a bulge-to-disc model, and 41 pure elliptical galaxies used as a comparison sample.

The basic description of the CALIFA survey, including the observational strategy and data reduction are explained in \citet{sanchez16}. In short, all galaxies were observed using the PMAS/PPaK instrument, which covers a hexagonal field of view (FoV) of 74 arcsec $\times$ 64 arcsec. The reconstructed datacubes guarantee a complete coverage of the FoV with a final spatial resolution of full width at half-maximum (FWHM) $\sim$ 2.5 arcsec, corresponding to $\sim$ 1 kpc at the average redshift of the survey \citep[][]{garciabenito15} and covering up to 2.5 $\times$ $r_{\rm e, g}$ (galaxy effective radius) for 90\% of the sample \citep{walcher14}. The wavelength range and spectroscopic resolution for the adopted V500 setup (3745-7500 \AA, $R \sim$ 850) are perfectly suited to explore the properties of the stellar populations and the ionised-gas emission lines.

\section{{\sc c2d}+{\sc Pipe3D} analysis of the sample}
\label{sec:c2d}

The spectro-photometric decomposition of the CALIFA ETGs into a bulge and disc was performed using {\sc c2d} \citep{mendezabreu19}. This new methodology allows us to separate the spectral contribution of each structural component providing an independent datacube for both the bulge and disc. For the sake of clarity, we will briefly describe the adopted procedure here.

The IFS datacubes from CALIFA can be understood as a set of quasi-monochromatic 2D images at different wavelengths. Therefore, we can apply a 2D photometric decomposition code to each of these images in order to isolate the contribution from both the bulge and disc. In {\sc c2d}, the photometric decomposition engine is provided by GASP2D \citep{mendezabreu08a,mendezabreu14}, a code that has been extensively tested on different galaxy samples with multiple structures \citep[][]{delorenzocaceres19a,delorenzocaceres19b}. We used a S\'ersic  \citep{sersic68} and an exponential \citep{freeman70} model to describe the light distribution of the bulge and disc components, respectively.  The  best fitted values of the intensities for each quasi-monochromatic image directly return the characteristic spectrum for each component. In addition, {\sc c2d} derives an independent datacube (with spatial and spectral information) for each component. To do this, for each quasi-monochromatic image, the $B/T$ and the disc-to-total ($D/T=1-B/T$) ratios are computed for each spaxel. Each fraction is then multiplied by the observed CALIFA datacube in that spaxel and wavelength. Rearranging these quasi-monochromatic images as a function of the wavelength produces the independent bulge and disc datacubes. Further details on the specific application of {\sc c2d} to CALIFA data are presented in \citet{mendezabreu19}.

\begin{figure*}
\begin{center}
\includegraphics[width=0.9\textwidth]{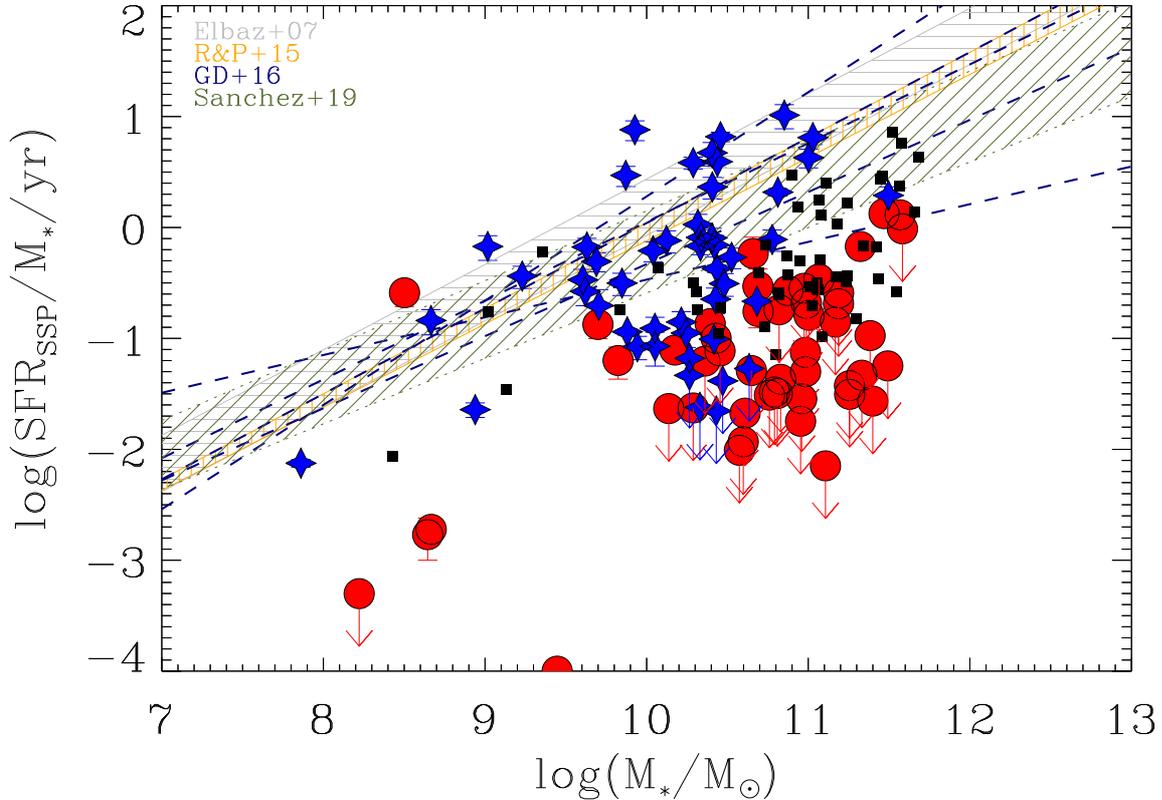}
\caption{Star formation rate (SFR) vs. stellar mass ($M_\star$) distribution for our sample of ETGs. The measurements for bulges and discs are shown with red circles and blue stars, respectively. Black squares represent measurements of the same ETGs, but for the galaxy as a whole. Yellow triangles display the position of the sample of elliptical galaxies described in the text. The best fit to the SFMS from \citet[Grey;][]{elbaz07}, \citet[Orange;][]{renzinipeng15}, \citet[Navy;][]{gonzalezdelgado16} and \citet[Green;][]{sanchez19} are also shown for comparison. Downward arrows mark where the measured SFR is an upper limit (see text for details).}
\label{fig:SFRmass}
\end{center}
\end{figure*}

The bulge and disc datacubes are then analysed using the {\sc Pipe3D} pipeline that uses the FIT3D tool \citep{sanchez16b}, which is specifically designed to extract the stellar population and ionised-gas properties from IFS data. In particular, {\sc Pipe3D} has been extensively tested on CALIFA data \citep[e.g.,][]{sanchez15,sanchez16a} and therefore the combination of {\sc c2d}+{\sc Pipe3D} represents the best strategy to derive the stellar and ionised-gas properties of our bulges and discs. The current implementation of {\sc Pipe3D} adopts the GSD156 library of simple stellar populations from \citet{cidfernandes13}, that comprises 156 templates covering 39 stellar ages (from 1 Myr to 13 Gyr) and four metallicities (Z/Z$_{\sun}$ = 0.2 dex, 0.4 dex, 1 dex, and 1.5 dex). The best-fit stellar-population model spectra to the galaxy continuum is subtracted from the original cube to create a gas-pure cube including the ionised-gas emission lines only. For this particular data set, we used the extracted flux intensities of the strong lines H$\alpha$ and H$\beta$. The intensity maps for each of these lines are corrected by dust attenuation, derived using the spaxel-to-spaxel H$\alpha$/H$\beta$ ratio. A canonical value of 2.86 is then assumed for this ratio \citep{osterbrock89}, and adopting a \citet{cardelli89} extinction law and $R_V$ = 3.1. We refer the reader to the presentation paper of {\sc Pipe3D} \citep{sanchez16a} for a detailed description of its application to CALIFA data.

\section{The star formation main sequence of ETGs}

Figure~\ref{fig:SFRmass} shows the SFR vs. $M_\star$ distribution for the bulges and discs of the 49 ETGs of our sample. The SFR was derived using the stellar population analysis described in Sect. \ref{sec:c2d} and the recipe given in \citet{gonzalezdelgado16}. In a nutshell, for any given galaxy we compute the spaxel-to-spaxel mass of stars younger than 32 Myrs. Then, by co-adding these masses over a given aperture we can estimate the average rate of star formation during this period. An extensive discussion about the optimal limiting age of the stellar population used to define the {\it recent} SFR is given in \citet{gonzalezdelgado16}. We tested the robustness of this assumption by repeating Figure~\ref{fig:SFRmass} using a limit of both 100 Myr and 200 Myr. We find that the results are qualitatively in agreement. Our stellar population approach to compute the SFR is different from the usual H$\alpha$ luminosity proxy used in the literature \citep{kennicutt98,catalantorrecilla15}. Nevertheless, it is most powerful for ETGs since it does not depend on the detection of emission lines. Fig.~\ref{fig:SFRmass} also shows the SFR vs. $M_\star$ relation for our comparison sample of ellipticals (see Sect.~\ref{sec:sample}) and for our ETGs without considering the bulge and disc components separately. It is worth noticing that, due to the optical coverage of the spectroscopic data, our SFR might be considered as upper limits for very low values of the SFR \citep[see discussion in][]{lopezfernandez18, bitsakis19}. This issue is more acute for our bulge SFR estimations, while it is almost negligible for the star-forming discs. However, this does not affect the main conclusions of this paper.

It is clear from Fig.~\ref{fig:SFRmass} that, whenever there is star formation in our ETGs, this is happening in the disc component and not in the bulge. We find that on average the specific SFR (sSFR) is smaller in bulges than in discs. The mean difference in logarithm accounts to at least 1.3 dex when integrated across the whole FoV of CALIFA. In addition, a qualitative comparison with literature values of the best fit to the SFMS at $z\sim0$ \citep{elbaz07,renzinipeng15,gonzalezdelgado16,sanchez19} shows that several discs of our ETG sample are compatible (or over) with the SFMS, whereas bulges occupy the region of the diagram defined by retired galaxies \citep{crocker11,canodiaz16,sanchez18}, or even lower SFR values for a given stellar mass. In particular, we find that 30 discs and only one bulge (NGC3773) show a SFR compatible (within 1$\sigma$) with the SFMS defined by \citet{sanchez19}. In fact, NGC3773 is not representative of the ETG classification since it hosts a strong starburst in its central regions, likely due to a recent merger, and most of the bulge light is dominated by young stellar populations. The elliptical galaxies in our sample are located, for a similar mass, in an intermediate position between bulges and discs, more similar to the position of the galaxies when they are consider as a whole (i.e., not separated into components).

\begin{figure}
\begin{center}
\includegraphics[width=0.49\textwidth]{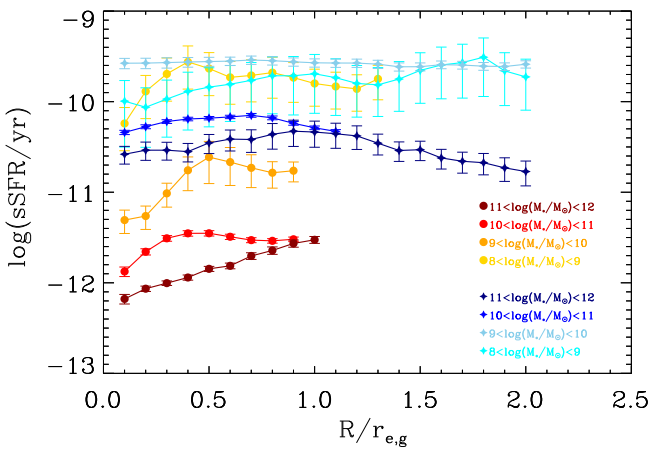}
\caption{sSFR radial profiles for the bulges (reddish colors) and discs (bluish colors) in our sample of ETGs. Different profiles represent the average distribution of the sSFR for the bulge/disc components within four different mass bins. The averaged radial profiles are normalised to the galaxy effective radius.}
\label{fig:ssfr_radial}
\end{center}
\end{figure}

The presence of residual star formation in the outer parts of ETGs has been previously reported in the literature from measurements of either emission lines \citep{sarzi06, gomes16} or atomic/molecular gas \citep{yildiz15,colombo18}. However, our spectro-photometric decomposition analysis demonstrates that the star formation in not only present in the outer regions of ETGs, but at all galactocentric radii when considering the disc component. This is shown in Fig.~\ref{fig:ssfr_radial} where the radial profiles of the sSFR for bulges and discs are displayed in different mass bins. Galaxy discs exhibit quite flat radial profiles for all different mass bins, indicating that the levels of star formation bringing these structures to follow the SFMS are maintained over the whole disc, and not just in their outer regions. The disc sSFR profiles are similar to those of Sc/Sd galaxies reported by \citet{gonzalezdelgado16}, supporting the presence of a disc within galaxy regions where the light is dominated by stars from the bulge. Galaxy bulges show different profiles with a drop at the center and a flat profile in their outer regions, possibly indicating an inside-out quenching \citep[e.g.,][]{sanchez18}. The expected trend between the sSFR and galaxy mass is clearly shown in Fig.~\ref{fig:ssfr_radial} and it holds for bulges and discs independently. We find that, for a given mass, bulges exhibit a lower sSFR than discs at all radii. Only the lowest-mass bulges ($8<\log{(M_{\star}/M_{\sun})}<9$) show sSFRs as high as discs of the same mass. However, our sample at these masses is small (4 galaxies) and it is dominated by the high values of NGC3773.  It is worth noticing that all the SFRs derived for the disc component in the C2D decomposition corresponds to reliable fractions of young stars in the {\sc Pipe3D} stellar analysis. \citet{bitsakis19} determined that the SFR estimations derived by {\sc Pipe3D} for galaxies with log(sSFR) $>$-11.5 yr$^{-1}$ are reliable ($\sim$ 0.15 dex), while values reported for lower sSFR (i.e., for RGs) should be considered just upper-limits (marked with downward arrows in Fig.~\ref{fig:SFRmass}). These limitations reinforce our results, since the separation between star-forming discs and retired bulges could be even larger.
 
The question that arises is: what is producing the lowered SFR in bulges with respect to discs? A possible explanation might be a different amount of cold molecular gas available to form stars in both components. We compute a first order estimation of the mass of molecular gas ($M_{\rm gas}$) in our bulges and discs based on the estimated dust attenuation from emission lines and the dust-to-gas ratio \citep[$M_{\rm gas,A_{V}}$;][]{brinchmann13, galbany17, sanchez18}. We find a similar $M_{\rm gas, A_{V}}$ as a function of the total galaxy stellar mass for the bulges and discs in our sample with averaged values $M_{\rm gas, A_{V}}$=8.9$\pm$0.8 and 8.3$\pm$0.8, respectively. However, these numbers are not fully representative of the sample since they are based on 17 bulges and 15 discs with detected emission lines. In order to have a estimation of $M_{\rm gas}$ for the whole sample we also use the average dust attenuation affecting the stellar populations ($A_{\rm ssp}$). We found a similar trend, but with larger uncertainties, for the bulges and discs with average values $M_{\rm gas, A_{ssp}}$=8.8$\pm$0.9 and 8.6$\pm$1.1, respectively. Further support to this result is provided by the CO observations obtained within the EDGE-CALIFA survey \citep{bolatto17}. Using these data, \citet{utomo17} and \citet{colombo18} found that the surface density of molecular gas is similar in the central and outer parts of galaxies and that it is nearly constant for all Hubble types. Therefore, even if a proper separation of the molecular gas into different galaxy components has not been attempted, since $M_{\rm gas, bulge} \sim M_{\rm gas, disc} $ and SFR$_{\rm bulge}$ $<$ SFR$_{\rm disc}$, it suggests a lower star formation efficiency in bulges than discs.

The results of this paper indicate that the absence/excess of molecular gas is not the main factor responsible for the different star formation in bulges and discs. We suggest that other processes such as gravitational stabilization against star formation (morphological quenching) might be driving the lower star formation efficiency in bulges.

\section{Conclusions}

We use the new algorithm ({\sc c2d}) to perform a bulge-to-disc spectro-photometric decomposition of a sample of 49 ETGs from the CALIFA DR3 survey. The stellar populations and ionised-gas analysis of the independent bulge and disc datacubes was then carried out using the {\sc Pipe3D} code. This unique combination allow us to study the SFR vs. $M_{\star}$ relation for bulges and discs to an unprecendented accuracy.

Despite ETGs are generally considered as pa\-ssi\-ve/retired galaxies, we found that whenever there is recent star formation this is present in the disc component and not in the bulge. Morever, these discs are compatible with the SFMS defined by star forming galaxies at $z\sim0$. On the contrary, bulges of ETGs show SFRs similar (or lower) than ellipticals of the same mass.

We find a flat radial profile of the sSFR for the discs, whereas they show a central drop for bulges. Therefore, the star formation in the discs of ETGs is not a phenomena only associated to their outer regions, but it also occurs in galaxy regions dominated by the bulge. This provides support for the presence of discs in the inner regions of galaxies.

We also demonstrate that bulges have systematically smaller sSFR (at all radii) than discs for the same mass. In addition, we estimate similar molecular gas masses for both bulges and discs in the ETG sample. Therefore, our results suggest that dynamical processes such as morphological quenching might be driving the mass growth of ETGs.

\section*{Acknowledgements}
We thank the referee for a constructive report which helped to improve the manuscript. JMA acknowledge support from the Spanish Ministerio de Economia y Competitividad (MINECO) by the grant AYA2017-83204-P. AdLC acknowledges support from grant AYA2016- 77237-C3-1-P from the Spanish Ministry of Economy and Competitiveness (MINECO). SFS thanks the projects ConaCyt CB-285080, FC-2016-01-1916 and PAPIIT IN100519. This paper is based on data from the Calar Alto Legacy Integral Field Area Survey, CALIFA, funded by the Spanish Ministery of Science under grant ICTS-2009-10, and the Centro Astron\'omico Hispano-Alem\'an. Based on observations collected at the Centro Astron\'omico Hispano Alem\'an (CAHA) at Calar Alto, operated jointly by the  Max-Planck Institut f\"ur Astronomie and the Instituto de Astrof\'isica de Andaluc\'ia.




\bibliographystyle{mnras}
\bibliography{reference}

\begin{thebibliography}{}
\makeatletter
\relax
\def\mn@urlcharsother{\let\do\@makeother \do\$\do\&\do\#\do\^\do\_\do\%\do\~}
\def\mn@doi{\begingroup\mn@urlcharsother \@ifnextchar [ {\mn@doi@}
  {\mn@doi@[]}}
\def\mn@doi@[#1]#2{\def\@tempa{#1}\ifx\@tempa\@empty \href
  {http://dx.doi.org/#2} {doi:#2}\else \href {http://dx.doi.org/#2} {#1}\fi
  \endgroup}
\def\mn@eprint#1#2{\mn@eprint@#1:#2::\@nil}
\def\mn@eprint@arXiv#1{\href {http://arxiv.org/abs/#1} {{\tt arXiv:#1}}}
\def\mn@eprint@dblp#1{\href {http://dblp.uni-trier.de/rec/bibtex/#1.xml}
  {dblp:#1}}
\def\mn@eprint@#1:#2:#3:#4\@nil{\def\@tempa {#1}\def\@tempb {#2}\def\@tempc
  {#3}\ifx \@tempc \@empty \let \@tempc \@tempb \let \@tempb \@tempa \fi \ifx
  \@tempb \@empty \def\@tempb {arXiv}\fi \@ifundefined
  {mn@eprint@\@tempb}{\@tempb:\@tempc}{\expandafter \expandafter \csname
  mn@eprint@\@tempb\endcsname \expandafter{\@tempc}}}

\bibitem[\protect\citeauthoryear{{Bitsakis} et~al.,}{{Bitsakis}
  et~al.}{2019}]{bitsakis19}
{Bitsakis} T.,  et~al., 2019, \mn@doi [\mnras] {10.1093/mnras/sty2857}, \href
  {http://adsabs.harvard.edu/abs/2019MNRAS.483..370B} {483, 370}

\bibitem[\protect\citeauthoryear{{Bluck}, {Mendel}, {Ellison}, {Moreno},
  {Simard}, {Patton}  \& {Starkenburg}}{{Bluck} et~al.}{2014}]{bluck14}
{Bluck} A.~F.~L.,  {Mendel} J.~T.,  {Ellison} S.~L.,  {Moreno} J.,  {Simard}
  L.,  {Patton} D.~R.,   {Starkenburg} E.,  2014, \mn@doi [\mnras]
  {10.1093/mnras/stu594}, \href
  {http://adsabs.harvard.edu/abs/2014MNRAS.441..599B} {441, 599}

\bibitem[\protect\citeauthoryear{{Bolatto} et~al.,}{{Bolatto}
  et~al.}{2017}]{bolatto17}
{Bolatto} A.~D.,  et~al., 2017, \mn@doi [\apj] {10.3847/1538-4357/aa86aa},
  \href {http://adsabs.harvard.edu/abs/2017ApJ...846..159B} {846, 159}

\bibitem[\protect\citeauthoryear{{Brinchmann}, {Charlot}, {White}, {Tremonti},
  {Kauffmann}, {Heckman}  \& {Brinkmann}}{{Brinchmann}
  et~al.}{2004}]{brinchmann04}
{Brinchmann} J.,  {Charlot} S.,  {White} S.~D.~M.,  {Tremonti} C.,  {Kauffmann}
  G.,  {Heckman} T.,   {Brinkmann} J.,  2004, \mn@doi [\mnras]
  {10.1111/j.1365-2966.2004.07881.x}, \href
  {http://adsabs.harvard.edu/abs/2004MNRAS.351.1151B} {351, 1151}

\bibitem[\protect\citeauthoryear{{Brinchmann}, {Charlot}, {Kauffmann},
  {Heckman}, {White}  \& {Tremonti}}{{Brinchmann} et~al.}{2013}]{brinchmann13}
{Brinchmann} J.,  {Charlot} S.,  {Kauffmann} G.,  {Heckman} T.,  {White}
  S.~D.~M.,   {Tremonti} C.,  2013, \mn@doi [\mnras] {10.1093/mnras/stt551},
  \href {http://adsabs.harvard.edu/abs/2013MNRAS.432.2112B} {432, 2112}

\bibitem[\protect\citeauthoryear{{Cano-D{\'{\i}}az} et~al.,}{{Cano-D{\'{\i}}az}
  et~al.}{2016}]{canodiaz16}
{Cano-D{\'{\i}}az} M.,  et~al., 2016, \mn@doi [\apjl]
  {10.3847/2041-8205/821/2/L26}, \href
  {http://adsabs.harvard.edu/abs/2016ApJ...821L..26C} {821, L26}

\bibitem[\protect\citeauthoryear{{Cappellari} et~al.,}{{Cappellari}
  et~al.}{2011}]{cappellari11}
{Cappellari} M.,  et~al., 2011, \mn@doi [\mnras]
  {10.1111/j.1365-2966.2011.18600.x}, \href
  {http://adsabs.harvard.edu/abs/2011MNRAS.416.1680C} {416, 1680}

\bibitem[\protect\citeauthoryear{{Cardelli}, {Clayton}  \& {Mathis}}{{Cardelli}
  et~al.}{1989}]{cardelli89}
{Cardelli} J.~A.,  {Clayton} G.~C.,   {Mathis} J.~S.,  1989, \mn@doi [\apj]
  {10.1086/167900}, \href {http://adsabs.harvard.edu/abs/1989ApJ...345..245C}
  {345, 245}

\bibitem[\protect\citeauthoryear{{Catal{\'a}n-Torrecilla}
  et~al.,}{{Catal{\'a}n-Torrecilla} et~al.}{2015}]{catalantorrecilla15}
{Catal{\'a}n-Torrecilla} C.,  et~al., 2015, \mn@doi [\aap]
  {10.1051/0004-6361/201526023}, \href
  {http://adsabs.harvard.edu/abs/2015A%26A...584A..87C} {584, A87}

\bibitem[\protect\citeauthoryear{{Cattaneo} et~al.,}{{Cattaneo}
  et~al.}{2009}]{cattaneo09}
{Cattaneo} A.,  et~al., 2009, \mn@doi [\nat] {10.1038/nature08135}, \href
  {http://adsabs.harvard.edu/abs/2009Natur.460..213C} {460, 213}

\bibitem[\protect\citeauthoryear{{Cid Fernandes} et~al.,}{{Cid Fernandes}
  et~al.}{2013}]{cidfernandes13}
{Cid Fernandes} R.,  et~al., 2013, \mn@doi [\aap]
  {10.1051/0004-6361/201220616}, \href
  {http://adsabs.harvard.edu/abs/2013A%26A...557A..86C} {557, A86}

\bibitem[\protect\citeauthoryear{{Colombo} et~al.,}{{Colombo}
  et~al.}{2018}]{colombo18}
{Colombo} D.,  et~al., 2018, \mn@doi [\mnras] {10.1093/mnras/stx3233}, \href
  {http://adsabs.harvard.edu/abs/2018MNRAS.475.1791C} {475, 1791}

\bibitem[\protect\citeauthoryear{{Cowie}, {Songaila}, {Hu}  \& {Cohen}}{{Cowie}
  et~al.}{1996}]{cowie96}
{Cowie} L.~L.,  {Songaila} A.,  {Hu} E.~M.,   {Cohen} J.~G.,  1996, \mn@doi
  [\aj] {10.1086/118058}, \href
  {http://adsabs.harvard.edu/abs/1996AJ....112..839C} {112, 839}

\bibitem[\protect\citeauthoryear{{Crocker}, {Bureau}, {Young}  \&
  {Combes}}{{Crocker} et~al.}{2011}]{crocker11}
{Crocker} A.~F.,  {Bureau} M.,  {Young} L.~M.,   {Combes} F.,  2011, \mn@doi
  [\mnras] {10.1111/j.1365-2966.2010.17537.x}, \href
  {http://adsabs.harvard.edu/abs/2011MNRAS.410.1197C} {410, 1197}

\bibitem[\protect\citeauthoryear{{Daddi} et~al.,}{{Daddi}
  et~al.}{2007}]{daddi07}
{Daddi} E.,  et~al., 2007, \mn@doi [\apj] {10.1086/521818}, \href
  {http://adsabs.harvard.edu/abs/2007ApJ...670..156D} {670, 156}

\bibitem[\protect\citeauthoryear{{Davis} et~al.,}{{Davis}
  et~al.}{2011}]{davis11}
{Davis} T.~A.,  et~al., 2011, \mn@doi [\mnras]
  {10.1111/j.1365-2966.2011.19355.x}, \href
  {http://adsabs.harvard.edu/abs/2011MNRAS.417..882D} {417, 882}

\bibitem[\protect\citeauthoryear{{Elbaz} et~al.,}{{Elbaz}
  et~al.}{2007}]{elbaz07}
{Elbaz} D.,  et~al., 2007, \mn@doi [\aap] {10.1051/0004-6361:20077525}, \href
  {http://adsabs.harvard.edu/abs/2007A%26A...468...33E} {468, 33}

\bibitem[\protect\citeauthoryear{{Ellison}, {Teimoorinia}, {Rosario}  \&
  {Mendel}}{{Ellison} et~al.}{2016}]{ellison16}
{Ellison} S.~L.,  {Teimoorinia} H.,  {Rosario} D.~J.,   {Mendel} J.~T.,  2016,
  \mn@doi [\mnras] {10.1093/mnrasl/slw012}, \href
  {http://cdsads.u-strasbg.fr/abs/2016MNRAS.458L..34E} {458, L34}

\bibitem[\protect\citeauthoryear{{Emsellem} et~al.,}{{Emsellem}
  et~al.}{2011}]{emsellem11}
{Emsellem} E.,  et~al., 2011, \mn@doi [\mnras]
  {10.1111/j.1365-2966.2011.18496.x}, \href
  {http://adsabs.harvard.edu/abs/2011MNRAS.414..888E} {414, 888}

\bibitem[\protect\citeauthoryear{{Feldmann}, {Carollo}, {Mayer}, {Renzini},
  {Lake}, {Quinn}, {Stinson}  \& {Yepes}}{{Feldmann} et~al.}{2010}]{feldmann10}
{Feldmann} R.,  {Carollo} C.~M.,  {Mayer} L.,  {Renzini} A.,  {Lake} G.,
  {Quinn} T.,  {Stinson} G.~S.,   {Yepes} G.,  2010, \mn@doi [\apj]
  {10.1088/0004-637X/709/1/218}, \href
  {http://adsabs.harvard.edu/abs/2010ApJ...709..218F} {709, 218}

\bibitem[\protect\citeauthoryear{{Freeman}}{{Freeman}}{1970}]{freeman70}
{Freeman} K.~C.,  1970, \mn@doi [\apj] {10.1086/150474}, \href
  {http://adsabs.harvard.edu/abs/1970ApJ...160..811F} {160, 811}

\bibitem[\protect\citeauthoryear{{Galbany} et~al.,}{{Galbany}
  et~al.}{2017}]{galbany17}
{Galbany} L.,  et~al., 2017, \mn@doi [\mnras] {10.1093/mnras/stx367}, \href
  {http://adsabs.harvard.edu/abs/2017MNRAS.468..628G} {468, 628}

\bibitem[\protect\citeauthoryear{{Garc{\'{\i}}a-Benito}
  et~al.,}{{Garc{\'{\i}}a-Benito} et~al.}{2015}]{garciabenito15}
{Garc{\'{\i}}a-Benito} R.,  et~al., 2015, \mn@doi [\aap]
  {10.1051/0004-6361/201425080}, \href
  {http://adsabs.harvard.edu/abs/2015A%26A...576A.135G} {576, A135}

\bibitem[\protect\citeauthoryear{{Gomes} et~al.,}{{Gomes}
  et~al.}{2016a}]{gomes16}
{Gomes} J.~M.,  et~al., 2016a, \mn@doi [\aap] {10.1051/0004-6361/201525974},
  \href {http://adsabs.harvard.edu/abs/2016A%26A...585A..92G} {585, A92}

\bibitem[\protect\citeauthoryear{{Gomes} et~al.,}{{Gomes}
  et~al.}{2016b}]{gomes16b}
{Gomes} J.~M.,  et~al., 2016b, \mn@doi [\aap] {10.1051/0004-6361/201525976},
  \href {http://adsabs.harvard.edu/abs/2016A%26A...588A..68G} {588, A68}

\bibitem[\protect\citeauthoryear{{Gonz{\'a}lez Delgado} et~al.,}{{Gonz{\'a}lez
  Delgado} et~al.}{2016}]{gonzalezdelgado16}
{Gonz{\'a}lez Delgado} R.~M.,  et~al., 2016, \mn@doi [\aap]
  {10.1051/0004-6361/201628174}, \href
  {http://adsabs.harvard.edu/abs/2016A%26A...590A..44G} {590, A44}

\bibitem[\protect\citeauthoryear{{Hopkins}, {Cox}, {Dutta}, {Hernquist},
  {Kormendy}  \& {Lauer}}{{Hopkins} et~al.}{2009}]{hopkins09}
{Hopkins} P.~F.,  {Cox} T.~J.,  {Dutta} S.~N.,  {Hernquist} L.,  {Kormendy} J.,
    {Lauer} T.~R.,  2009, \mn@doi [\apjs] {10.1088/0067-0049/181/1/135}, \href
  {http://adsabs.harvard.edu/abs/2009ApJS..181..135H} {181, 135}

\bibitem[\protect\citeauthoryear{{Jeong}, {Bureau}, {Yi}, {Krajnovi{\'c}}  \&
  {Davies}}{{Jeong} et~al.}{2007}]{jeong07}
{Jeong} H.,  {Bureau} M.,  {Yi} S.~K.,  {Krajnovi{\'c}} D.,   {Davies} R.~L.,
  2007, \mn@doi [\mnras] {10.1111/j.1365-2966.2007.11535.x}, \href
  {https://ui.adsabs.harvard.edu/abs/2007MNRAS.376.1021J} {376, 1021}

\bibitem[\protect\citeauthoryear{{Kaviraj} et~al.,}{{Kaviraj}
  et~al.}{2007}]{kaviraj07}
{Kaviraj} S.,  et~al., 2007, \mn@doi [\apjs] {10.1086/516633}, \href
  {http://adsabs.harvard.edu/abs/2007ApJS..173..619K} {173, 619}

\bibitem[\protect\citeauthoryear{{Kennicutt}}{{Kennicutt}}{1998}]{kennicutt98}
{Kennicutt} Jr. R.~C.,  1998, \mn@doi [\araa] {10.1146/annurev.astro.36.1.189},
  \href {http://cdsads.u-strasbg.fr/abs/1998ARA%26A..36..189K} {36, 189}

\bibitem[\protect\citeauthoryear{{Kormendy} \& {Bender}}{{Kormendy} \&
  {Bender}}{2012}]{kormendybender12}
{Kormendy} J.,  {Bender} R.,  2012, \mn@doi [\apjs]
  {10.1088/0067-0049/198/1/2}, \href
  {http://adsabs.harvard.edu/abs/2012ApJS..198....2K} {198, 2}

\bibitem[\protect\citeauthoryear{{Laurikainen}, {Salo}, {Buta}  \&
  {Knapen}}{{Laurikainen} et~al.}{2011}]{laurikainen11}
{Laurikainen} E.,  {Salo} H.,  {Buta} R.,   {Knapen} J.~H.,  2011, \mn@doi
  [\mnras] {10.1111/j.1365-2966.2011.19283.x}, \href
  {http://adsabs.harvard.edu/abs/2011MNRAS.418.1452L} {418, 1452}

\bibitem[\protect\citeauthoryear{{L{\'o}pez Fern{\'a}ndez} et~al.,}{{L{\'o}pez
  Fern{\'a}ndez} et~al.}{2018}]{lopezfernandez18}
{L{\'o}pez Fern{\'a}ndez} R.,  et~al., 2018, \mn@doi [\aap]
  {10.1051/0004-6361/201732358}, \href
  {http://adsabs.harvard.edu/abs/2018A%26A...615A..27L} {615, A27}

\bibitem[\protect\citeauthoryear{{Martig}, {Bournaud}, {Teyssier}  \&
  {Dekel}}{{Martig} et~al.}{2009}]{martig09}
{Martig} M.,  {Bournaud} F.,  {Teyssier} R.,   {Dekel} A.,  2009, \mn@doi
  [\apj] {10.1088/0004-637X/707/1/250}, \href
  {http://adsabs.harvard.edu/abs/2009ApJ...707..250M} {707, 250}

\bibitem[\protect\citeauthoryear{{Martig} et~al.,}{{Martig}
  et~al.}{2013}]{martig13}
{Martig} M.,  et~al., 2013, \mn@doi [\mnras] {10.1093/mnras/sts594}, \href
  {http://adsabs.harvard.edu/abs/2013MNRAS.432.1914M} {432, 1914}

\bibitem[\protect\citeauthoryear{{M{\'e}ndez-Abreu}, {Aguerri}, {Corsini}  \&
  {Simonneau}}{{M{\'e}ndez-Abreu} et~al.}{2008}]{mendezabreu08a}
{M{\'e}ndez-Abreu} J.,  {Aguerri} J.~A.~L.,  {Corsini} E.~M.,   {Simonneau} E.,
   2008, \mn@doi [\aap] {10.1051/0004-6361:20078089}, \href
  {http://adsabs.harvard.edu/abs/2008A%26A...478..353M} {478, 353}

\bibitem[\protect\citeauthoryear{{M{\'e}ndez-Abreu}, {Debattista}, {Corsini}
  \& {Aguerri}}{{M{\'e}ndez-Abreu} et~al.}{2014}]{mendezabreu14}
{M{\'e}ndez-Abreu} J.,  {Debattista} V.~P.,  {Corsini} E.~M.,   {Aguerri}
  J.~A.~L.,  2014, \mn@doi [\aap] {10.1051/0004-6361/201423955}, \href
  {http://adsabs.harvard.edu/abs/2014A%26A...572A..25M} {572, A25}

\bibitem[\protect\citeauthoryear{{M{\'e}ndez-Abreu} et~al.,}{{M{\'e}ndez-Abreu}
  et~al.}{2017}]{mendezabreu17}
{M{\'e}ndez-Abreu} J.,  et~al., 2017, \mn@doi [\aap]
  {10.1051/0004-6361/201629525}, \href
  {http://adsabs.harvard.edu/abs/2017A%26A...598A..32M} {598, A32}

\bibitem[\protect\citeauthoryear{{M{\'e}ndez-Abreu} et~al.,}{{M{\'e}ndez-Abreu}
  et~al.}{2018}]{mendezabreu18}
{M{\'e}ndez-Abreu} J.,  et~al., 2018, \mn@doi [\mnras] {10.1093/mnras/stx2804},
  \href {http://adsabs.harvard.edu/abs/2018MNRAS.474.1307M} {474, 1307}

\bibitem[\protect\citeauthoryear{{M{\'e}ndez-Abreu}, {S{\'a}nchez}  \& {de
  Lorenzo-C{\'a}ceres}}{{M{\'e}ndez-Abreu} et~al.}{2019}]{mendezabreu19}
{M{\'e}ndez-Abreu} J.,  {S{\'a}nchez} S.~F.,   {de Lorenzo-C{\'a}ceres} A.,
  2019, \mn@doi [\mnras] {10.1093/mnras/stz276}, \href
  {http://adsabs.harvard.edu/abs/2019MNRAS.484.4298M} {484, 4298}

\bibitem[\protect\citeauthoryear{{Moffett}, {Kannappan}, {Baker}  \&
  {Laine}}{{Moffett} et~al.}{2012}]{moffett12}
{Moffett} A.~J.,  {Kannappan} S.~J.,  {Baker} A.~J.,   {Laine} S.,  2012,
  \mn@doi [\apj] {10.1088/0004-637X/745/1/34}, \href
  {http://adsabs.harvard.edu/abs/2012ApJ...745...34M} {745, 34}

\bibitem[\protect\citeauthoryear{{Noeske} et~al.,}{{Noeske}
  et~al.}{2007}]{noeske07}
{Noeske} K.~G.,  et~al., 2007, \mn@doi [\apjl] {10.1086/517926}, \href
  {http://adsabs.harvard.edu/abs/2007ApJ...660L..43N} {660, L43}

\bibitem[\protect\citeauthoryear{{Osterbrock}}{{Osterbrock}}{1989}]{osterbrock89}
{Osterbrock} D.~E.,  1989, {Astrophysics of gaseous nebulae and active galactic
  nuclei}

\bibitem[\protect\citeauthoryear{{Renzini} \& {Peng}}{{Renzini} \&
  {Peng}}{2015}]{renzinipeng15}
{Renzini} A.,  {Peng} Y.-j.,  2015, \mn@doi [\apjl]
  {10.1088/2041-8205/801/2/L29}, \href
  {http://adsabs.harvard.edu/abs/2015ApJ...801L..29R} {801, L29}

\bibitem[\protect\citeauthoryear{{Saintonge} et~al.,}{{Saintonge}
  et~al.}{2012}]{saintonge12}
{Saintonge} A.,  et~al., 2012, \mn@doi [\apj] {10.1088/0004-637X/758/2/73},
  \href {http://cdsads.u-strasbg.fr/abs/2012ApJ...758...73S} {758, 73}

\bibitem[\protect\citeauthoryear{{Salim} \& {Rich}}{{Salim} \&
  {Rich}}{2010}]{salimrich10}
{Salim} S.,  {Rich} R.~M.,  2010, \mn@doi [\apjl]
  {10.1088/2041-8205/714/2/L290}, \href
  {http://adsabs.harvard.edu/abs/2010ApJ...714L.290S} {714, L290}

\bibitem[\protect\citeauthoryear{{Salim} et~al.,}{{Salim}
  et~al.}{2007}]{salim07}
{Salim} S.,  et~al., 2007, \mn@doi [\apjs] {10.1086/519218}, \href
  {http://adsabs.harvard.edu/abs/2007ApJS..173..267S} {173, 267}

\bibitem[\protect\citeauthoryear{{S{\'a}nchez} et~al.,}{{S{\'a}nchez}
  et~al.}{2012}]{sanchez12}
{S{\'a}nchez} S.~F.,  et~al., 2012, \mn@doi [\aap]
  {10.1051/0004-6361/201117353}, \href
  {http://adsabs.harvard.edu/abs/2012A%26A...538A...8S} {538, A8}

\bibitem[\protect\citeauthoryear{{S{\'a}nchez} et~al.,}{{S{\'a}nchez}
  et~al.}{2015}]{sanchez15}
{S{\'a}nchez} S.~F.,  et~al., 2015, \mn@doi [\aap]
  {10.1051/0004-6361/201424873}, \href
  {http://adsabs.harvard.edu/abs/2015A%26A...574A..47S} {574, A47}

\bibitem[\protect\citeauthoryear{{S{\'a}nchez} et~al.,}{{S{\'a}nchez}
  et~al.}{2016a}]{sanchez16b}
{S{\'a}nchez} S.~F.,  et~al., 2016a, \rmxaa, \href
  {http://adsabs.harvard.edu/abs/2016RMxAA..52...21S} {52, 21}

\bibitem[\protect\citeauthoryear{{S{\'a}nchez} et~al.,}{{S{\'a}nchez}
  et~al.}{2016b}]{sanchez16a}
{S{\'a}nchez} S.~F.,  et~al., 2016b, \rmxaa, \href
  {http://adsabs.harvard.edu/abs/2016RMxAA..52..171S} {52, 171}

\bibitem[\protect\citeauthoryear{{S{\'a}nchez} et~al.,}{{S{\'a}nchez}
  et~al.}{2016c}]{sanchez16}
{S{\'a}nchez} S.~F.,  et~al., 2016c, \mn@doi [\aap]
  {10.1051/0004-6361/201628661}, \href
  {http://adsabs.harvard.edu/abs/2016A%26A...594A..36S} {594, A36}

\bibitem[\protect\citeauthoryear{{S{\'a}nchez} et~al.,}{{S{\'a}nchez}
  et~al.}{2018}]{sanchez18}
{S{\'a}nchez} S.~F.,  et~al., 2018, \rmxaa, \href
  {http://cdsads.u-strasbg.fr/abs/2018RMxAA..54..217S} {54, 217}

\bibitem[\protect\citeauthoryear{{S{\'a}nchez} et~al.,}{{S{\'a}nchez}
  et~al.}{2019}]{sanchez19}
{S{\'a}nchez} S.~F.,  et~al., 2019, \mn@doi [\mnras] {10.1093/mnras/sty2730},
  \href {http://cdsads.u-strasbg.fr/abs/2019MNRAS.482.1557S} {482, 1557}

\bibitem[\protect\citeauthoryear{{Sarzi} et~al.,}{{Sarzi}
  et~al.}{2006}]{sarzi06}
{Sarzi} M.,  et~al., 2006, \mn@doi [\mnras] {10.1111/j.1365-2966.2005.09839.x},
  \href {http://adsabs.harvard.edu/abs/2006MNRAS.366.1151S} {366, 1151}

\bibitem[\protect\citeauthoryear{{Sarzi} et~al.,}{{Sarzi}
  et~al.}{2010}]{sarzi10}
{Sarzi} M.,  et~al., 2010, \mn@doi [\mnras] {10.1111/j.1365-2966.2009.16039.x},
  \href {http://adsabs.harvard.edu/abs/2010MNRAS.402.2187S} {402, 2187}

\bibitem[\protect\citeauthoryear{{Schawinski}, {Thomas}, {Sarzi}, {Maraston},
  {Kaviraj}, {Joo}, {Yi}  \& {Silk}}{{Schawinski} et~al.}{2007}]{schawinski07}
{Schawinski} K.,  {Thomas} D.,  {Sarzi} M.,  {Maraston} C.,  {Kaviraj} S.,
  {Joo} S.-J.,  {Yi} S.~K.,   {Silk} J.,  2007, \mn@doi [\mnras]
  {10.1111/j.1365-2966.2007.12487.x}, \href
  {http://adsabs.harvard.edu/abs/2007MNRAS.382.1415S} {382, 1415}

\bibitem[\protect\citeauthoryear{{Schawinski}, {Virani}, {Simmons}, {Urry},
  {Treister}, {Kaviraj}  \& {Kushkuley}}{{Schawinski}
  et~al.}{2009}]{schawinski09}
{Schawinski} K.,  {Virani} S.,  {Simmons} B.,  {Urry} C.~M.,  {Treister} E.,
  {Kaviraj} S.,   {Kushkuley} B.,  2009, \mn@doi [\apjl]
  {10.1088/0004-637X/692/1/L19}, \href
  {http://adsabs.harvard.edu/abs/2009ApJ...692L..19S} {692, L19}

\bibitem[\protect\citeauthoryear{{Schiminovich} et~al.,}{{Schiminovich}
  et~al.}{2007}]{schiminovich07}
{Schiminovich} D.,  et~al., 2007, \mn@doi [\apjs] {10.1086/524659}, \href
  {http://cdsads.u-strasbg.fr/abs/2007ApJS..173..315S} {173, 315}

\bibitem[\protect\citeauthoryear{{Scott} \& {Kaviraj}}{{Scott} \&
  {Kaviraj}}{2014}]{scottkaviraj14}
{Scott} C.,  {Kaviraj} S.,  2014, \mn@doi [\mnras] {10.1093/mnras/stt2014},
  \href {http://cdsads.u-strasbg.fr/abs/2014MNRAS.437.2137S} {437, 2137}

\bibitem[\protect\citeauthoryear{{Serra} et~al.,}{{Serra}
  et~al.}{2012}]{serra12}
{Serra} P.,  et~al., 2012, \mn@doi [\mnras] {10.1111/j.1365-2966.2012.20219.x},
  \href {http://adsabs.harvard.edu/abs/2012MNRAS.422.1835S} {422, 1835}

\bibitem[\protect\citeauthoryear{{S\'ersic}}{{S\'ersic}}{1968}]{sersic68}
{S\'ersic} J.~L.,  1968, {Atlas de galaxias australes}

\bibitem[\protect\citeauthoryear{{Shapiro} et~al.,}{{Shapiro}
  et~al.}{2010}]{shapiro10}
{Shapiro} K.~L.,  et~al., 2010, \mn@doi [\mnras]
  {10.1111/j.1365-2966.2009.16111.x}, \href
  {http://adsabs.harvard.edu/abs/2010MNRAS.402.2140S} {402, 2140}

\bibitem[\protect\citeauthoryear{{Speagle}, {Steinhardt}, {Capak}  \&
  {Silverman}}{{Speagle} et~al.}{2014}]{speagle14}
{Speagle} J.~S.,  {Steinhardt} C.~L.,  {Capak} P.~L.,   {Silverman} J.~D.,
  2014, \mn@doi [\apjs] {10.1088/0067-0049/214/2/15}, \href
  {http://cdsads.u-strasbg.fr/abs/2014ApJS..214...15S} {214, 15}

\bibitem[\protect\citeauthoryear{{Tacconi} et~al.,}{{Tacconi}
  et~al.}{2013}]{tacconi13}
{Tacconi} L.~J.,  et~al., 2013, \mn@doi [\apj] {10.1088/0004-637X/768/1/74},
  \href {http://cdsads.u-strasbg.fr/abs/2013ApJ...768...74T} {768, 74}

\bibitem[\protect\citeauthoryear{{Thomas}, {Maraston}, {Schawinski}, {Sarzi}
  \& {Silk}}{{Thomas} et~al.}{2010}]{thomas10}
{Thomas} D.,  {Maraston} C.,  {Schawinski} K.,  {Sarzi} M.,   {Silk} J.,  2010,
  \mn@doi [\mnras] {10.1111/j.1365-2966.2010.16427.x}, \href
  {http://adsabs.harvard.edu/abs/2010MNRAS.404.1775T} {404, 1775}

\bibitem[\protect\citeauthoryear{{Utomo} et~al.,}{{Utomo}
  et~al.}{2017}]{utomo17}
{Utomo} D.,  et~al., 2017, \mn@doi [\apj] {10.3847/1538-4357/aa88c0}, \href
  {http://adsabs.harvard.edu/abs/2017ApJ...849...26U} {849, 26}

\bibitem[\protect\citeauthoryear{{Walcher} et~al.,}{{Walcher}
  et~al.}{2014}]{walcher14}
{Walcher} C.~J.,  et~al., 2014, \mn@doi [\aap] {10.1051/0004-6361/201424198},
  \href {http://adsabs.harvard.edu/abs/2014A%26A...569A...1W} {569, A1}

\bibitem[\protect\citeauthoryear{{Willett} et~al.,}{{Willett}
  et~al.}{2015}]{willett15}
{Willett} K.~W.,  et~al., 2015, \mn@doi [\mnras] {10.1093/mnras/stv307}, \href
  {http://cdsads.u-strasbg.fr/abs/2015MNRAS.449..820W} {449, 820}

\bibitem[\protect\citeauthoryear{{Wuyts} et~al.,}{{Wuyts}
  et~al.}{2011}]{wuyts11}
{Wuyts} S.,  et~al., 2011, \mn@doi [\apj] {10.1088/0004-637X/742/2/96}, \href
  {http://cdsads.u-strasbg.fr/abs/2011ApJ...742...96W} {742, 96}

\bibitem[\protect\citeauthoryear{{Y{\i}ld{\i}z}, {Serra}, {Oosterloo},
  {Peletier}, {Morganti}, {Duc}, {Cuillandre}  \& {Karabal}}{{Y{\i}ld{\i}z}
  et~al.}{2015}]{yildiz15}
{Y{\i}ld{\i}z} M.~K.,  {Serra} P.,  {Oosterloo} T.~A.,  {Peletier} R.~F.,
  {Morganti} R.,  {Duc} P.-A.,  {Cuillandre} J.-C.,   {Karabal} E.,  2015,
  \mn@doi [\mnras] {10.1093/mnras/stv992}, \href
  {http://adsabs.harvard.edu/abs/2015MNRAS.451..103Y} {451, 103}

\bibitem[\protect\citeauthoryear{{Young} et~al.,}{{Young}
  et~al.}{2011}]{young11}
{Young} L.~M.,  et~al., 2011, \mn@doi [\mnras]
  {10.1111/j.1365-2966.2011.18561.x}, \href
  {http://adsabs.harvard.edu/abs/2011MNRAS.414..940Y} {414, 940}

\bibitem[\protect\citeauthoryear{{de Lorenzo-C{\'a}ceres}, {M{\'e}ndez-Abreu},
  {Thorne}  \& {Costantin}}{{de Lorenzo-C{\'a}ceres}
  et~al.}{2019a}]{delorenzocaceres19a}
{de Lorenzo-C{\'a}ceres} A.,  {M{\'e}ndez-Abreu} J.,  {Thorne} B.,
  {Costantin} L.,  2019a, \mn@doi [\mnras] {10.1093/mnras/sty3520}, \href
  {http://adsabs.harvard.edu/abs/2019MNRAS.484..665D} {484, 665}

\bibitem[\protect\citeauthoryear{{de Lorenzo-C{\'a}ceres} et~al.,}{{de
  Lorenzo-C{\'a}ceres} et~al.}{2019b}]{delorenzocaceres19b}
{de Lorenzo-C{\'a}ceres} A.,  et~al., 2019b, \mn@doi [\mnras]
  {10.1093/mnras/stz221}, \href
  {http://adsabs.harvard.edu/abs/2019MNRAS.484.5296D} {484, 5296}

\makeatother
\end{thebibliography}





\bsp	
\label{lastpage}
\end{document}